\documentstyle{europhys}

\def\And{{\rm and\ }}

\def\stars{\bigskip\centerline{***}\medskip}

\newif\ifboo \boofalse

\def\Review#1{\boofalse{\it #1},}
\def\Name#1{{\sc #1},}
\def\Vol#1{\ifboo Vol. {\bf #1}\else{\bf #1}\fi}
\def\Year#1{\ifboo #1\else(#1)\fi}

\def\Page#1{\ifboo {\rm p. #1}\else{\rm #1}\fi}

\begin{document}
\euro{xx}{y}{1-$\infty$}{1999}
\Date{28 June 1999}

\shorttitle{xxx.... }

\title{Magnetic relaxation in hard type-II superconductors}

\author{Mahesh Chandran}

\institute{Department of Condensed Matter Physics and Material Sciences,\\ 
Tata Institute of Fundamental Research, Colaba, Mumbai, India 400 005.}

\rec{zz June 1999}{in final form xxx}

\pacs{
\Pacs{74}{60Ge}{Flux pinning, flux creep, and flux line lattice dynamics}
\Pacs{74}{60Jg}{Critical currents}
}
\maketitle

\begin{abstract}
Magnetic relaxation in a type-II superconductor is simulated for a range of temperatures $T$ in a simple model of 2D Josephson junction array (JJA) with finite screening. The high-$T$ phase, that is characterised by a single time scale $\tau_{\alpha}$, crosses over to an intermediate phase at a lower temperature $T_{cr}$ wherein a second time scale $\tau_{\beta}\ll\!\tau_{\alpha}$ emerges. The relaxation in the time window set by $\tau_{\beta}$ follows power law which is attributed to  self-organization of the magnetic flux during relaxation. Consequently, for  $T<T_{cr}$, a transition from super-critical (current density $J>J_{c}$) to sub-critical ($J<J_{c}$) state separated by an intermediate state with frozen dynamics is observed. Both $\tau_{\alpha}$ and $\tau_{\beta}$ diverges at $T_{sc}<T_{cr}$, marking the transition into a state with true persistent current. 
\end{abstract}

\section{Introduction} 
In a hard type-II superconductor, flux-creep over energy barrier at a finite $T$ leads to magnetic relaxation over long time scales. The flux creep, occurring due to thermally activated hopping of the flux lines, tend to reduce the local field gradient $\frac{dB}{dx}$, and hence the current density $J$. In analyzing relaxation measurements, the initial magnetic field distribution is assumed to be that of Bean's critical state \cite{bean} wherein $J$ is replaced by the critical current density $J_c$. If the relaxation is close to $J_c$, the magnetisation decay $M(t)$ is then theoretically known to be logarithmic, as is the case in most low-$T_c$ superconductors \cite{kim}. In this case, the effective pinning potential $U(J)\sim U_{0}(J-J_{c})$ is a good approximation.

In high-$T_c$ superconductors (HTSC), the large thermal energy available leads to rapid decay of $M(t)$. Experimentally the relaxation is observed to be non-logarithmic over several decades in time in these materials \cite{yesh}. This is interpreted as arising from a non-linear $U(J)$. 
The vortex-glass theory \cite{fisher} and collective-creep theory \cite{fiegel}, which predicts a low temperature true superconducting state with finite $J_{c}$, expects a pinning potential of the form $U(J)=U_{0}[(J_{c}/J)^{\mu}-1]$ which diverges in the limit $J\rightarrow 0$. An important experimental observation in HTSC is an apparent universal value of the normalized relaxation rate $S(T)=\frac{-1}{M}\frac{dM}{d\ln t}$ around 0.02-0.035 over a wide range of $T$ \cite{malz}. For the $U(J)$ given above, such a small $T$-independent $S$ requires $\mu >2$ which is beyond the range of existing theoretical models \cite{yesh}. Experiments have also indicated power law decay in HTSC \cite{maley} which requires a logarithmically diverging $U(J)=U_{o}\ln(J_{c}/J)$ \cite{zeld,vinokur}. This form of $U(J)$ cannot account for the experimentally observed plateau in $S(T)$.

It becomes then important to identify the relaxation behaviour of the magnetisation in a type-II superconductor in presence of a uniform background of pinning centers. Towards this end, we simulate the time decay of remanent magnetisation $M(t)$ at finite $T$ in a simple model of 2D Josephson junction array (JJA). The magnetisation in this model is studied by including the screening currents through the inductance $L_{\bf R,R'}$ between the cells. The underlying discrete lattice of junctions provides an energy barrier for the vortex motion within the array \cite{lobb} and is the source of vortex pinning in JJA. The behaviour of JJA with screening is parameterized by $\lambda_{J}^{2}=\frac{\Phi_{0}}{2\pi L_{0}I_{c}}$, where $I_{c}$ is the critical current of the junction, $L_{0}$ is the self-inductance of the cell, and $\Phi_{0}$ is the quantum of flux. Detailed simulation at $T=0$ have shown that for $\lambda_{J}^{2}<1$, the magnetic response of this model is similar to a continuum hard type-II superconductor 
\cite{maj,mpc,phil,mah}. 
In this paper, we show that the thermo-remanent relaxation of $M(t)$ of JJA captures essential features of magnetic relaxation in HTSC. 
At a crossover temperature $T_{cr}$, a new time scale $\tau_{\beta}$ emerges in an intermediate time window which along with the characteristic time scale $\tau_{\alpha}$ for long time behaviour governs the relaxation. The characteristic time scales $\tau_{\alpha}$ and $\tau_{\beta}$ diverges as a power law at a temperature $T_{sc}$ at which $M(\tau)\rightarrow M_{0}\neq 0$ as $\tau\rightarrow\infty$.

\section{The Model}
We consider a 2D array of superconducting islands forming a homogeneous square lattice of $N\times N$ unit cells in the $x\!-\!y$ plane. Tunneling of the macroscopic wavefunction $\Psi({\bf r})=\psi\exp[i\varphi({\bf r})]$ across neighbouring islands lead to Josephson coupling between them. In presence of an applied magnetic flux $\Phi_{ext}$ (per cell) along $\hat{z}$ direction, the junction behaviour is fully described by dynamics of the gauge-invariant phase difference 
$\phi_{{\bf r},\delta}\!=\!\varphi({\bf r})-\!\varphi({\bf r+\delta})\!-\!\frac{2\pi}{\Phi_{0}}\int_{\bf r}^{\bf r+\delta}{\bf A}\cdot d{\bf l}$ between neighbouring islands. Here, $\delta$ is a unit vector, and ${\bf A}$ is the vector potential corresponding to the total magnetic field. The inset of Fig.1 shows a schematic array of size $N=5$ along with variable $\phi_{{\bf r},x}$ and $\phi_{{\bf r},y}$.

The magnetic response of JJA is set by the induced flux $\Phi_{ind}({\bf R})$ due to screening currents which is modeled by considering the geometrical inductance matrix $L$ of the array \cite{phil}. This allows us to write the induced flux in a cell at ${\bf R}$ as $\Phi_{ind}({\bf R})=\sum_{\bf R'} L({\bf R,R'})I({\bf R'})$ where $I({\bf R'})$ is the cell current at ${\bf R'}$ and $L({\bf R,R'})$ is the mutual inductance between the cells at ${\bf R}$ and ${\bf R'}$. In order to ease the prohibitive computation cost involved when mutual inductance is considered \cite{phil,mah}, we consider the induced flux only due to self-inductance $L_{0}=L({\bf R,R})$ of the cell. This approximation is equivalent to the case of a long (ideally infinite) superconductor parallel to an applied field for which demagnetisation factor $N_{D}=0$. The total flux at {\bf R} then can be written as $\Phi_{\bf R}=\Phi_{ext}+L_{0}I_{\bf R}$ (the cell co-ordinates are used as subscripts). The cell current is only a convenient variable for introducing $\Phi_{ind}$ as the current through the junction is $I_{{\bf r},\delta}=I_{\bf R}-I_{\bf R-\delta}$ where {\bf r} is the junction common to cells at {\bf R} and ${\bf R-\delta}$. Since, $I_{\bf R}$ is divergence-less, Kirchoff's law is automatically satisfied at each node of the lattice.

The dynamical variable $\phi_{\bf r}$ (subscript $\delta$ is implicit) is related to the cell current $I_{\bf R}$ through the flux-quantisation condition
\begin{equation}
\sum_{{\bf r}\in{\bf R}}\phi_{\bf r}\; =\; -2\pi\frac{\Phi_{\bf R}}{\Phi_{0}}= -2\pi\frac{\Phi_{ext}}{\Phi_{0}}  - \frac{2\pi}{\Phi_{0}}L_{0}I_{\bf R},
\end{equation}
where the summation is taken around the cell in anti-clockwise direction. The equation of motion for $\phi_{{\bf r},\delta}$ then follows from 
\begin{equation}
\frac{d\phi_{\bf r}}{dt} = -\Gamma \frac{\delta{\cal H}}{\delta\phi_{\bf r}}, 
\end{equation}
where the Hamiltonian ${\cal H}$ of the system is 
\begin{equation}
{\cal H}= \sum_{{\bf r}} E_{J}(1-\cos\phi_{{\bf r}})\; + \; \frac{1}{2}\sum_{{\bf R}}L_{0} I_{{\bf R}}^{2},
\end{equation}
where $\Gamma=(\frac{\Phi_{0}}{2\pi})^{2}\frac{1}{R}$ with $R$ and $I_{c}$ as the normal state resistance and critical current of the junction respectively. In the above equation, the first term is summed over all bonds and represent Josephson coupling energy for the junction, whereas the second term is the magnetic field energy due to screening currents and the summation is over all cells in the array.

The above equations can be written compactly by introducing the matrix {\sf M} for the lattice curl operation \cite{phil}. Eq.(1) then becomes ${\sf M}\phi\!=\!-2\pi f\!-\!\frac{2\pi}{\Phi_{0}}L_{0}I_{m}$ where $\phi$ and $I_{m}$ are the column vectors formed by $\phi_{\bf r}$ and $I_{\bf R}$ respectively, and $f=\Phi_{ext}/\Phi_{0}$. Also, the current through the junction $I_{b}={\sf M}^{T}I_{m}$, where ${\sf M}^{T}$ is transpose of the matrix {\sf M}. For finite temperature simulation, we couple the heat bath through the noise current in the shunt resistor $R$ across the junction. Langevin equation for the array then takes a simple form  
\begin{eqnarray}
\frac{d\phi}{d\tau} & \;= & \;{\sf M}^{T}{I}_{m}\;-\; \sin \phi\;\;+\;\; X(\tau),
\nonumber \\
{\sf M}\phi & \;= & \;-2\pi f\;-\;\frac{1}{\lambda^{2}_{J}}{I}_{m}.
\end{eqnarray}
Here, the current $I_{m}$ is scaled by the critical current $I_{c}$ of the junction. The dimensionless time $\tau=\frac{2\pi RI_{c}}{\Phi_{0}}t$, and $\lambda^{2}_{J}$ has been defined earlier. The random term $X(\tau)$ has zero mean and white noise correlation
\begin{equation}
\langle X_{\bf r}(\tau) \rangle\!=\!0, \;\;{\rm and}\;\; \langle X_{\bf r}(\tau)X_{\bf r'}(\tau') \rangle \!= \!2T\delta(\tau-\tau')\delta_{\bf r,r'},
\end{equation}
where $T$ is the temperature of the bath (in units of $I_{c}\Phi_{0}/2\pi k_{B}$). In this unit, the phase coherence between neighbouring islands is established at $T=1$ and is the superconducting transition. The set of equations in (4) is solved self-consistently at each time step with free-end boundary condition. For simplicity, $I_{c}$ is assumed to be independent of $f$ and $T$ (the $f$ dependence does not show any qualitative change in the results presented below). The magnetisation $M$ is obtained as $M=(1/N^{2})\sum_{\bf R}(\Phi_{\bf R}/\Phi_{0})-f$.

The simulations were performed for $N=16(\equiv 256$ cells) with time step $\Delta\tau=0.05$. The results presented below are for $\lambda_{J}^{2}=0.1$. Note that for $\lambda_{J}^{2}<1$, each cell accommodates more than a quantum of flux. The moderate size of the array allowed long time to be reached in the relaxation that is essential at low temperatures (the longest relaxation time run was of $\approx 5\times10^{6}\tau\approx 10^{8}$ iterations). The field cooling is done by quenching from a high temperature (typically $T=2$) to the temperature $T$ under consideration, followed by annealing for $5\!\times\! 10^{3}\!-\!8\!\times\! 10^{3}\tau$ before switching off the applied field at $\tau=0$ (thermoremanent magnetisation). We have also carried out simulation for slow cooling which shows no qualitative difference from that obtained after a sudden quench, though a small but discernible dependence on the cooling rate is observed. We defer such effects and further details to a future paper, and present here results which brings out generic features.

\section{Results and discussions}

The $M(\tau)$ is shown in Fig.1 over 7 decades of time after cooling to different temperatures in an applied field $f=5$ (curves for intermediate $T$ is not shown in order to avoid over crowding of the figure). The curves are scaled by $M(\tau\!=\!0)$. From the curves, two distinct temperatures $T_{cr}\approx 0.26$ and $T_{sc}\approx 0.04$ can be identified at which the dynamical behaviour changes remarkably. For $T>T_{cr}$, $M(\tau)$ can be characterized by a single time scale $\tau_{\alpha}$. At $T=T_{cr}$, $M(\tau)$ develops a kink in an intermediate time window which at lower temperatures evolve into a plateau. The $\tau_{\alpha}$ now characterizes the long time behaviour of $M(\tau)$. On the plateau, the magnetisation $M(\tau)=M_{0}$ is $T$-independent. The temporary freezing of dynamics indicates emergence of a new time scale, represented by $\tau_{\beta}$, which governs the relaxation for $\tau<<\tau_{\alpha}$. Both $\tau_{\alpha}$ and $\tau_{\beta}$ increases rapidly at lower temperatures, and at $T_{sc}$, $M(\tau)$ freezes asymptotically to $M_{0}$ for the time probed in simulation (note that for $T_{sc}<T$, $M(\tau)$ becomes zero in the limit $\tau\rightarrow\infty$). The temperature $T_{sc}$ thus marks the transition into a state with true persistent current. For $T<T_{sc}$, $M(\tau)$ relaxes towards $M_{0}$.

Also shown in the inset of Fig.1 is the magnetic susceptibility $\chi=M/f$ under field cooled (FC) and zero-field cooled (ZFC) conditions (for ZFC, $f$ is applied at $T=0$ and $\chi$ is calculated while increasing $T$). For $T>T_{cr}$, the $\chi_{FC}$ and $\chi_{ZFC}$ are equal and $M(T)$ is reversible. At $T_{cr}$, the difference $\mid\chi_{FC}-\chi_{ZFC}\mid >0$, and the $\chi_{ZFC}$ below it depends strongly upon the thermal history \cite{mah1}. Similar behaviour in bulk superconductors allow us to identify $T_{cr}$ as the irreversibility temperature. Though $\chi_{FC}$ does not show any change at $T_{sc}$, $\chi_{ZFC}$ is $T$ independent below it. The $T$ independent value of $\chi_{ZFC}\approx 0.6$ gives $M_{ZFC}\approx\! 3.0$ which equals remanent magnetisation $M_{0}$ obtained from the relaxation for $T\leq T_{sc}$ and is also found to hold for other values of $f$. This equality between the $M_{0}$ and $M_{ZFC}$ can be understood by invoking the relation $M_{rem}\!(H)\!=\!M_{FC}\!(H)\!-\!M_{ZFC}\!(H)$ ($H$ is the applied field) which is experimentally known to be valid in bulk type-II superconductors \cite{malz1}. In case of strong flux pinning, $M_{FC}\approx 0$ and $M_{rem}\approx -M_{ZFC}$, which is the case observed here. Validity of the above relation is also an evidence for strong flux pinning in JJA arising due to discrete underlying lattice of junctions. Further, we analyze the relaxation occurring on different time scales.

The long time decay of $M(\tau)$, which we term as $\alpha$-relaxation, fits Kohlrausch-Williams-Watt (KWW) law $M(\tau)\!=\!\exp[-(\frac{\tau}{\tau_{\alpha}})^{\alpha}]$ over the temperature range $T>T_{sc}$ probed in the simulation. 
The $\tau_{\alpha}(T)$ shows 6 orders of increase between $T_{sc}$ and $T_{cr}$ implying that the system falls out of equilibrium on cooling through it. Though $\tau_{\alpha}(T)$ appears to fit Arrhenius law $\tau_{\alpha}(T)\sim e^{A/T}$ (see Fig.2 inset (a)), notably, a distinct power law is observed at low temperatures as shown in Fig.2 where $\tau_{\alpha}^{-1}$ is plotted against $T-T_{sc}$. Fit to $\tau_{\alpha}(T)=\tau_{0}\mid T-T_{sc}\mid^{-\gamma}$ gives $T_{sc}=0.045, \gamma=3.74$ for $f=5$, and $T_{sc}=0.05, \gamma=4.735$ for $f=2$ which is also included in the figure. The value of $T_{sc}$ obtained from the fit is not very different from that at which the relaxation is frozen asymptotically at $M_{0}$ but it must be treated with considerable reserve. It is probable that the actual value may be lower than that obtained here as the time scale probed at low temperatures put severe constraint on obtaining $\tau_{\alpha}$. In Fig.2, the deviation from linearity occurs at $T\approx 0.24$ which is close to $T_{cr}$ below which the relaxation shows a temporary frozen state. The $T$ dependence of the exponent $\alpha$ is also shown in Fig.2 (inset (b)). Due to small value of $\alpha$ at low temperatures, the $M(\tau)$ can be fit to $\log\tau$ over 1-2 decades in time, and is the regime where thermally activated flux-creep theory can be applied \cite{kim}.

Appearance of a new time scale $\tau_{\beta}$ for $T<T_{cr}$ leads to two-step relaxation process : an initial part {\em towards} the plateau during $\tau_{0}\!\ll\!\tau\!\ll\!\tau_{\beta}$, and a later part {\em away} from the plateau in the interval  $\tau_{\beta}\!\ll\!\tau\!\ll\!\tau_{\alpha}$ which at later time develops into $\alpha$-relaxation. Here, $\tau_{0}\approx 20\tau$ is of the same order which appears in the power law fit for $\tau_{\alpha}$. We refer the relaxation during $\tau_{0}\ll\tau\ll\tau_{\alpha}$ as the $\beta$-relaxation regime due to apparent qualitative similarity with the $\beta$-relaxation seen in supercooled liquids \cite{gotze}. For $T_{sc}<T$, the relaxation during $\tau_{\beta}\!\ll\!\tau\!\ll\!\tau_{\alpha}$ fits a power law $M(\tau)-M_{0} \sim -c(\tau/\tau_{\beta})^b$ over 3 decades with exponent $b$ independent of $T$. This is shown in Fig.3 where $M(\tau)$ (for $f=5$) for different temperatures can be scaled on to a master curve when plotted against $\tau/\tau_{\alpha}(T)$. The thick line is the power law fit with $b=0.85$ and holds for $\tau/\tau_{\alpha}(T)\ll 1$ as evident from the fit. The scaling exponent $b$ allow us to obtain the $T$ dependence of $\tau_{\beta}(T)$ which is shown in Fig.3 (inset). $\tau_{\beta}(T)$ fits to a power law of the form $\tau_{\beta}=\tau_{0}\mid T-T_{sc}\mid^{-\psi}$ with $\psi=1.58$ and $T_{sc}=0.033$ ($\tau_{0}$ is of the same order as that obtained for $\tau_{\alpha}$). Since the $\beta$-relaxation at later time evolves into $\alpha$-relaxation, divergence in $\tau_{\beta}$ and $\tau_{\alpha}$ must occur at the same temperature. We attribute the small difference seen here to uncertainties in obtaining $\tau_{\alpha}$. Nevertheless, divergence of $\tau_{\beta}$ and $\tau_{\alpha}$ at $T_{sc}$ is significant as it implies absence of flux-creep (hence, divergent $U(J)$) as $\tau\rightarrow\infty$, and is consistent with the idea of true superconducting state below a transition temperature.

For $\tau_{0}\!\ll\!\tau\!\ll\!\tau_{\beta}$, the relaxation occur {\em towards} the plateau, and for $T<T_{sc}$ is asymptotically frozen on it. The relaxation can again be fit to a power law $M(\tau)-M_{0}\sim c'(\tau/\tau_{\beta})^{-a}$. The exponent $a$ is dependent on $T$ as shown in Fig.4. Also shown in the figure is $S(T)$ close to plateau which attains a $T$ independent value $\approx 0.076$ (on the plateau, $S(T)$ is an order of magnitude lower than this). This value is of the same order as that observed in HTSC for which it falls in the range $0.02-0.035$ \cite{malz}. A plateau is $S(T)$ coupled with non-logarithmic decay of $M(\tau)$ over 4 decades is significant in view of similar experimental observation in HTSC. To understand the processes involved during the relaxation, the spatial distribution of $\Phi_{\bf R}$ in different time windows is obtained and is shown in the inset of Fig.4. The slope of the flux profile $(\partial\Phi/\partial R_{x})\propto J$ is observed to be $T$-independent on the plateau and equals the slope ($\propto J_{c}$) in the remanent state at $T=0$ (note that $I_c$ of the junction is assumed to be $T$-independent in the simulation). At $T=0$, such a state have been shown to be the self-organized Bean's critical state for a hard superconductor \cite{mpc}.

Thus, across the $\beta$-relaxation regime, the current density relaxes from the super-critical state $J\!>\!J_{c}$ towards the sub-critical state $J\!<\!J_{c}$. The power law behaviour in this regime can be attributed to the self-organization of magnetic flux around the critical current density $J=J_{c}$. The self-organization of vortices is driven by inter-vortex repulsion and vortex-pinning center attraction alone \cite{pla}, and is independent of the temperature. This explains the plateau in $S(T)$ in Fig.4, and is also consistent with the apparent universality of $S(T)$ repeatedly observed in HTSC \cite{malz}. Moreover, in HTSC the critical current density $J_c$ is 1-2 orders of magnitude less than that of low-$T_c$ superconductors over a wide region of $H-T$ phase diagram. Since the relaxation towards $M_{0}\propto J_c$ is a power law, the time required to set up critical state in HTSC is much longer compared to low-$T_c$ superconductors. This implies that relaxation measurement in HTSC is influenced by the self-organization for much longer time and could be the source of non-logarithmic behaviour. It is important to note that the $\log t$ relaxation due to thermal activation becomes a dominant process only in the sub-critical state.

The results can be summarized by a $\tau\!-\! T$ plot for the magnetic relaxation in hard type-II superconductors which is shown in Fig.5. 
Emergence of a new time scale at a temperature at which history dependence sets in defines dynamically the irreversibility temperature for hard superconductors. The regime $T_{cr}<T$ is analogous to the vortex liquid (VL) phase with critical current density $J_{c}=0$. With decreasing $T$ below $T_{cr}$, the flux motion becomes rapidly viscous as evident from the power law increase in $\tau_{\alpha}$. The overall behaviour is analogous to the ``supercooled'' state in glass formers 
\cite{gotze}. In these system, relaxation of the correlation in density fluctuation shows scaling behaviour in the $\beta$-relaxation regime \cite{kob} which is remarkably similar to Fig.3 here. Relaxation data in HTSC need to be reanalyzed to observe the $T$ dependence of the characteristic time scale across the irreversibility temperature and glass transition temperature as observed here. Also, in fabricated JJA, SQUID parameter $\beta_{J}=\lambda_{J}^{-2}$ of 30 have been achieved \cite{arauro} which falls within the parameter range in which the simulation results can be applied.

In conclusion, simulation of the magnetic relaxation in a model of hard superconductor shows emergence of a new time scale below the irreversibility temperature. Self-organization of the magnetic flux around $J_{c}$ in this time scale leads to power law decay of the magnetisation. Divergence of time scales at a transition temperature leads to a state with finite remanent magnetisation (and hence, persistent current). The temperature dependence of the normalized relaxation rate is in good agreement with experiments on HTSC.

\stars

{\bf Acknowledgements :}
The author is grateful to D. Dhar and A. K. Grover for useful discussions and critical reading of the manuscript.

\section{Figure Captions}

\newcounter{bean}

\begin{list}%
{Fig.\arabic{bean}}{\usecounter{bean}}

\item Relaxation of the remanent magnetisation $M(\tau)$ on $\log\tau$ scale for $\lambda^{2}_{J}=0.1,f=5$ and array size $16\times 16$. The curves (from left to right) are for temperature $T$=0.9, 0.6, 0.4, 0.3, 0.24, and 0.20 to 0.02 in steps of 0.02. Inset: Plot of $\chi(T)$ for $f=5$ in FC and ZFC state. The irreversibility temperature is also the crossover temperature $T_{cr}$ (marked in the inset) at which $M(\tau)$ develops a kink. Also shown is a typical $5\times 5$ array with $\phi_{{\bf r},x}, \phi_{{\bf r},y}$ and $I_{\bf R}$ for a single cell.

\item Log-log plot of $\tau_{\alpha}^{-1}(T)$ against $T-T_{sc}$ for all values of $T$ for which the simulation was performed. The symbols $\bigtriangledown$ and $\circ$ are for $f=2, f=5$, respectively. The errors are less than the symbol size. The fitting parameter $T_{sc}$ and $\gamma$ is also given in the plot. Inset: (a) Arrhenius plot $\tau_{\alpha}(T)$ vs $T^{-1}$. (b) The stretching parameter $\alpha$ as a function of $T$.

\item $M(\tau)$ vs rescaled time $\tau/\tau_{\alpha}(T)$ for $0.05\leq T\leq 0.13$ (in steps of 0.01) in the $\beta$-relaxation regime. The thick dashed line is a fit to $M(\tau)\sim\tau^{-b}$ with $b=0.85$. Inset : $\tau_{\beta}$ as a function of $T-T_{sc}$ for $0.05\leq T\leq 0.14$ obtained by fitting the relaxation away from the plateau. The full line is a fit to $\tau_{\beta}\sim (T-T_{sc})^{-\psi}$ with $\psi =1.58\pm 0.2$ and $T_{sc}=0.033$.

\item The exponent $a\propto\! S$ (normalized relaxation rate) for $T\leq 0.12$. Also shown is the $S$ obtained from $M(\tau)$ around the plateau in Fig.1. Inset : The flux distribution $\Phi_{\bf R}/\Phi_{0}$ across a central row of cells in the array for various $\tau$ at $T=0.07$ for $f=5$.

\item The $\tau-T$ plot showing various regimes obtained from the simulation. The dotted line is obtained by interpolating $\tau_{\beta}(T)$ to $T_{cr}$. A and B are the power law regimes with exponents $a$ and $b$, respectively.
\end{list}

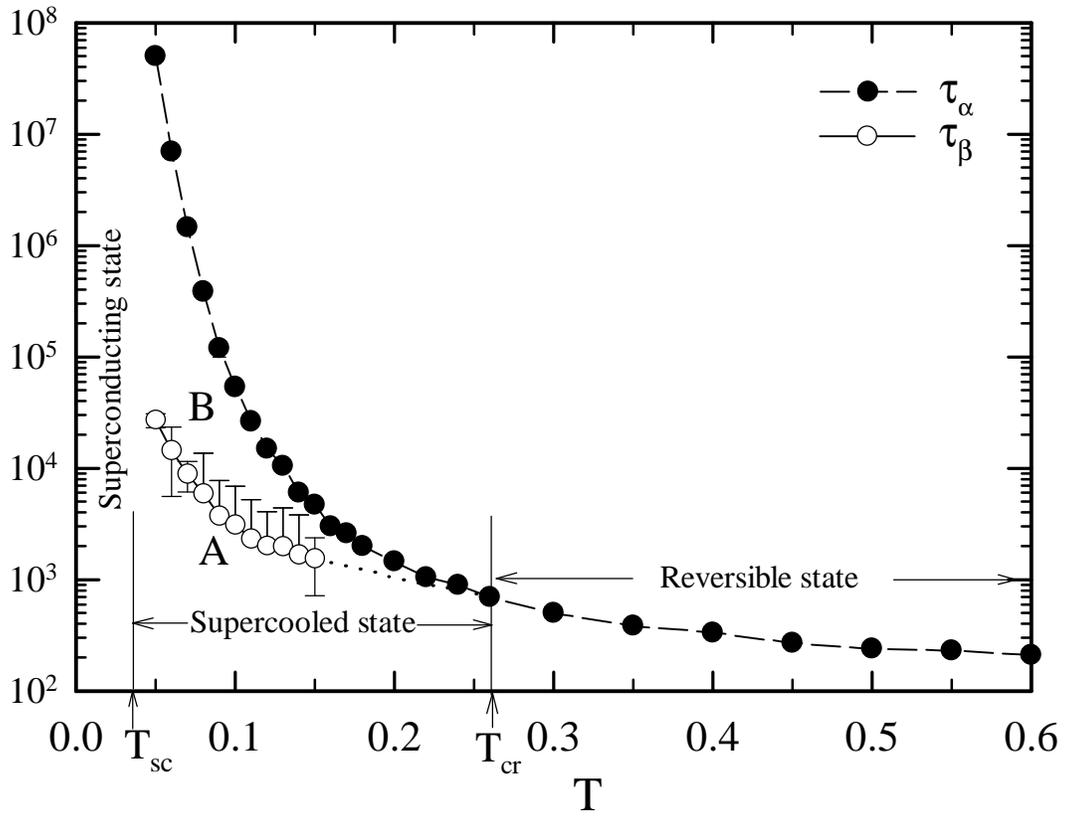
\end{document}